\documentclass[%
 reprint,
 amsmath,amssymb,
 aps,
]{revtex4-1}

\usepackage{graphicx}
\usepackage{dcolumn}
\usepackage{bm}

\begin{document}

\preprint{APS/123-QED}

\title{Past of a particle in an entangled state}

\author{Dilip Paneru}
 \email{dilippaneru@gmail.com}
 \affiliation{%
 	Department of Electronics and Computer Engineering, Central Campus, Pulchowk, Institute of Engineering, Tribhuvan University, Nepal
 }%

\author{Eliahu Cohen}
 \email{eliahu.cohen@bristol.ac.uk}
 \affiliation{%
 	H.H. Wills Physics Laboratory, University of Bristol,\\ Tyndall Avenue, Bristol, BS8 1TL, U.K 	
 }%

\date{\today}

\begin{abstract}
Vaidman has proposed a controversial criterion for determining the past of a single quantum particle based on the ``weak trace'' it leaves. We here consider more general examples of entangled systems and analyze the past of single, as well as pairs of entangled pre- and postselected particles. Systems with non-trivial time evolution are also analyzed. We argue that in these cases, examining only the single-particle weak trace provides information which is insufficient for understanding the system as a whole. We therefore suggest to examine, alongside with the past of single particles, also the past of pairs, triplets and eventually the entire system, including higher-order, multipartite traces in the analysis. This resonates with a recently proposed top-down approach by Aharonov, Cohen and Tollaksen for understanding the structure of correlations in pre- and postselected systems.
\end{abstract}

\maketitle

\section{\label{intro}Introduction}
The Two-State-Vector-Formalism (TSVF) \cite{tsvf,CohenAharonov} is a time-symmetric framework within quantum mechanics that is particularly suited for the analysis of ensembles which are both pre- and postselected. When augmented with the complementary concept of weak measurements \cite{weak_meas} this description allows us to pose and then answer questions which are otherwise impossible. The minimally disturbing coupling between the measured system and measuring device in a weak measurement scheme, induces only a negligible change in the measured system, which therefore allows to probe it without practically altering it.

One seminal example of the above, is inferring the $\sigma_{\pi/4}=(\sigma_x+\sigma_y)/\sqrt{2}$ component of a spin which was preselected with $|\psi\rangle=|\sigma_x=+1\rangle$ and postselected with $|\phi\rangle=|\sigma_y=+1\rangle$, using the weak value $\langle \sigma_{\pi/4} \rangle_w = \langle \phi | \sigma_{\pi/4} |\psi \rangle/\langle \phi | \psi \rangle=\sqrt{2}$, while maintaining the preselected state, as well as the probability of post-selection. This so-called ``anomalous weak value'' assigns an intriguing history to the pre- and postselected system. Another example, on which we will focus in this paper, is determining the past of quantum particles using the weak trace they leave \cite{pastofaparticle}.

Vaidman has suggested an intriguing criterion for determining the past of a single pre- and postselected particle \cite{pastofaparticle,tracing_the_past} based on the overlap between the forward- and backward-evolving wavefunctions. It was proposed that the particle was present in places where it left a trace, that is a non-zero weak value, which can be detected by weak measurements. The deflection of the weakly coupled measurement pointer is proportional to the weak value of the corresponding projection operator, i.e. to the weak trace of the particle. Vaidman's criterion has drawn much attention and various aspects of his approach are currently under debate \cite{zubiary,replytozubiary,nikolaev,replytonikolaev,weakmeasurementcriteria,hashmi,replytohashmi,griffiths,replytogriffiths,griffiths1,weakvalueandlogic,Englert}.

 We argue that Vaidman's criterion, when applied to single-particle states and product states yields unambiguous and meaningful answers regarding the past of the particle. However, the criterion has to be broadened to accommodate the cases where  particles are preselected and/or postselected in entangled states (see also \cite{ACbook}). As the single-particle traces are quantum mechanically correlated through multipartite entanglement, the information obtained only from the single-particle traces is incomplete. When considered individually, the multi-particle contributions  reveal further details about the past. Due to a special property of dichotomic operators \cite{dichotomic}, both the single-particle and multi-particle presences can be validated through strong measurement whenever the weak value of the corresponding dichotomic operator coincides with one of its eigenvalues. Additionally, the trace of single-particle projection operators and the trace of  multi-particle projection operators, if considered separately, can sometimes give apparently contradictory answers to the past question. This is a consequence of the well-known failure of  ``product rule for weak values'' \cite{failureproduct}, and Svensson in \cite{weakvalueandlogic} has argued that this leads to logical inconsistencies when compared to strong measurements. However, rather than considering them as contradictory, we suggest considering both the single-particle traces and the higher order ones to obtain a more complete information about the past of the quantum system. We thus address Svensson's concerns and try to refine the logic behind weak values. We will also show that the significance of the multi-particle weak traces accords well with a recently suggested top-down logical structure in quantum mechanics \cite{top-down}, according to which there is a sense in which many-body correlations are more fundamental than single-particle properties or few-body correlations.

In the following sections we analyze different examples of entangled systems using the weak value based time-symmetric approach and discuss the past of quantum particles in general multipartite scenarios.

\section{\label{challenges}Challenges in entangled systems}

Vaidman's criterion is motivated by whether the particle has any physical effect on an object, weakly coupled to it, which resides in the immediate proximity of its path. Since it is concerned with observable interactions, amenable to laboratory experiments, the criterion is physically meaningful. Furthermore, whenever a non-zero weak trace of a projection operator is detected in a certain path, there is a non-zero probability of observing the particle in a strong measurement. In fact, as mentioned above, when the weak value of the projection operator is one or zero, the presence/absence can be verified by a strong measurement as well \cite{dichotomic}. The predictions obtained from the weak trace criterion regarding the past of a particle were also verified in several experiments \cite{askingphoton,askingphoton2}. Furthermore, it was shown how to measure sequential weak values revealing the particle's past at several instances in time \cite{Pia}. Using the weak value criterion to determine the past has an additional advantage. If we use only the standard forward-evolving wavefunction, we have to keep track of a lot of information. Instead, considering both the forward- and backward-evolving wavefunctions, we can discard the excess information, and work with only the absolutely necessary information concerning both pre- and postselected states.

The method of determining where was a particle in the past by using weak trace of projection operators gives unambiguous (yet still controversial \cite{zubiary,replytozubiary,nikolaev,replytonikolaev,weakmeasurementcriteria,hashmi,replytohashmi,griffiths,replytogriffiths,griffiths1,weakvalueandlogic,Englert}) answer for single-partice/product states. However, in the case of entangled states the trace of single-particle projection operators provides us only with part of the whole picture. This is an inevitable consequence of quantum nonlocality. To motivate our alternative approach employing correlations between single-particle pasts, we shall begin with two examples which underscore the importance of multipartite correlations while considering the past of a particle in entangled states.

\subsection{\label{hardy}Hardy's Paradox}
Hardy's thought experiment \cite{Hardy,Hardy2} analyzes an electron-positron pair in a setup with two Mach-Zehnder interferometers overlapping in one corner. When the electron and positron are simultaneously present in the overlapping arm they annihilate each other.  The interferometers are tuned such that the electron entering the first interferometer always arrives at $ C^{-}$ and the positron entering the second always arrives at $ C^{+}$. In the case when both particles simultaneously enter the setup and no-annihilation occurs the (entangled) state of the particles is
\begin{equation}
 |\psi\rangle = \frac{1}{\sqrt{3}}\left(|O^{+}\rangle|NO^{-}\rangle + |NO^{+}\rangle|O^{-}\rangle + |NO^{+}\rangle|NO^{-}\rangle
 \right).
\end{equation}
If we consider only the cases where $D^{+}$ and $D^{-}$ click, then the postselected state is,
\begin{equation}
\resizebox{1\hsize}{!}{$|\phi\rangle = \frac{1}{2}\left(|O^{+}\rangle|O^{-}\rangle-|O^{+}\rangle|NO^{-}\rangle-|NO^{+}\rangle|O^{-}\rangle + |NO^{+}\rangle|NO^{-}\rangle\right).$}
\end{equation}
The weak values of various projection operators measured at some intermediate time are
\begin{equation}
P_{O^{+}O^{-}}=0,
\end{equation}
and
\begin{equation}
P_{O^{+}}= 1 = P_{O^{-}}.
\end{equation}

\begin{figure}
	\includegraphics[width=\linewidth]{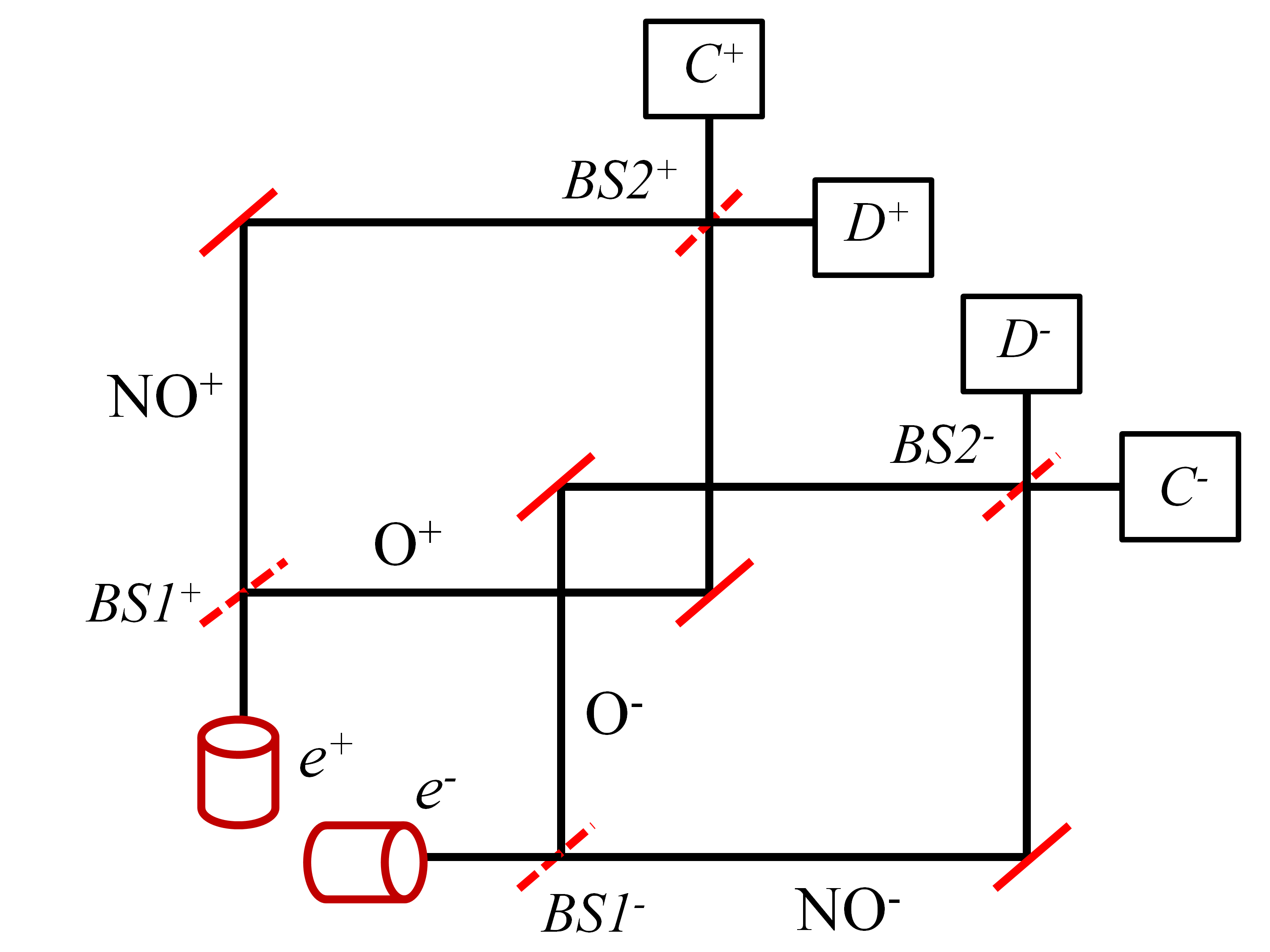}
	\caption{Hardy's thought experiment.}
	\label{}
\end{figure}

The weak values of single-particle projection operators tell us that the electron and positron are individually present in the overlapping arms. This has to be true, otherwise the detectors could not have detected the particles at $ D^{+}$ and $ D^{-} $. However, the electron and positron were not {\it simultaneously} present in the overlapping arm, as the weak value of the projection operator of the pair is zero. This has to be true because if they were simultaneously present there, then annihilation would have occurred.

 Next we consider a complementary example in which the single-particle traces do not indicate the presence of the particles, but the trace of the pair does.

\subsection{\label{simpleex}A simple example}
Let us consider two particles present in two boxes denoted by 1 and 2, pre- and postselected as follows
\begin{equation}
|\psi\rangle = \frac{1}{\sqrt{3}}\left(|11\rangle - |12\rangle + |22\rangle\right),
\end{equation}
\begin{equation}
|\phi\rangle = \frac{1}{\sqrt{3}}\left(|11\rangle + |12\rangle + |22\rangle\right).
\end{equation}
Let us define the projection operators as
\begin{equation}
\begin{aligned}
P_1^{(1)}&=|1\rangle \langle1| \otimes I, & P_2^{(2)}&=I\otimes|2\rangle \langle 2| \\
P_{11}&= |11\rangle \langle11|,  & P_{12}&=|12\rangle \langle12| \\ P_{21}&=|21\rangle \langle21|, & P_{22}&=|22\rangle \langle22|.
\end{aligned}
\end{equation}
The weak values of the projection operators are
\begin{equation}
\begin{aligned}
\langle P_1^{(1)}\rangle_{w}&=0, & \langle P_2^{(2)}\rangle_{w}&=0,
& \langle P_{11}\rangle_{w}&=1 \\ \langle P_{12}\rangle_{w}&=-1, &
\langle P_{22}\rangle_{w}&=1.
\end{aligned}
\end{equation}
We can see that the weak values of the single-particle projection operators, $P_{1}^{(1)}$ and $P_{2}^{(2)}$ are zero, but the weak values of the projection operators of the pairs, $P_{11}$, $P_{12}$ and $P_{22}$, are not zero. In other words when we look for the particles individually we do not find them, but when looked for together they are found there. If we consider only the single-particle traces (as done by Vaidman), we would conclude that the first particle is not in the box 1 and similarly that the second particle is not at box 2.
However the weak traces corresponding to pairs tell an additional story. We have,
\begin{eqnarray}
\begin{aligned}
\langle P_{1}^{(1)}\rangle_{w} = \langle P_{11}\rangle_{w} + \langle P_{12}\rangle_{w} = 1 - 1 = 0, \\
\langle P_{2}^{(2)}\rangle_{w} = \langle P_{12}\rangle_{w} + \langle P_{22}\rangle_{w} = 1 - 1 = 0.
\end{aligned}
\end{eqnarray}
	
The traces of the pair cancel each other to give a zero single-particle trace, but individually they are non-zero.
The boxes are therefore empty according to Vaidman's criterion, but in fact, they are full of correlations. This may suggest a subtle presence of the first particle in box 1 and the second's in box 2. Moreover, if we measure the presence of the two particles together in the first box, i.e. $P_{11}$ in the first box (or $P_{22}$ in the second box) strongly, then we are guaranteed to find them there due to the aforementioned property of dichotomic operators \cite{dichotomic}.

As suggested in a recent paper \cite{top-down}, the complete set of higher order correlations is sufficient to determine all the lower order ones, but the opposite is not true. The many-particle correlations cannot be deduced from the lower order correlations between fewer particles.  To elaborate further on that, let us reexamine the example above: Suppose we have weak values of the complete set of two-particle projection operators $P_{11}$, $P_{12}$, $P_{21}$, and $P_{22}$. From this set we can easily obtain the single-particle weak values. For instance,
$\langle P^{(1)}_{1}\rangle_{w}=\langle P_{11}\rangle_{w}+\langle P_{12}\rangle_{w}$ and
$\langle P^{(1)}_{2}\rangle_{w}=\langle P_{21}\rangle_{w}+\langle P_{22}\rangle_{w}$.
But the converse is not true. If we know all the single-particle weak values we cannot calculate the two-particle correlations.
For instance $P_{11} = P^{(1)}_{1} P^{(2)}_{1}$, but by virtue of the failure of product rule \cite{failureproduct} for weak values, we have $\langle P_{11}\rangle_{w} \neq \langle P^{(1)}_{1}\rangle_{w} \langle P^{(2)}_{1}\rangle_{w}$. Thus, we cannot obtain the nonlocal multipartite correlations from the set of single-particle correlations.
This discussion also further highlights the role of multipartite traces  in determining the past of the particle.

\subsection*{\label{secondary presence} Comparison with Vaidman's notion of ``secondary presence''}
Vaidman has also introduced  a notion of secondary presence \cite{tracing_the_past} in regions where the first-order weak trace vanishes, yet the forward- and backward-evolving wavefunctions still have a non-zero (yet small) overlap. The local interactions in such places can change the weak trace in the overlap region, but they do not alter the probability of postselection.
\begin{figure}
	\includegraphics[width=\linewidth]{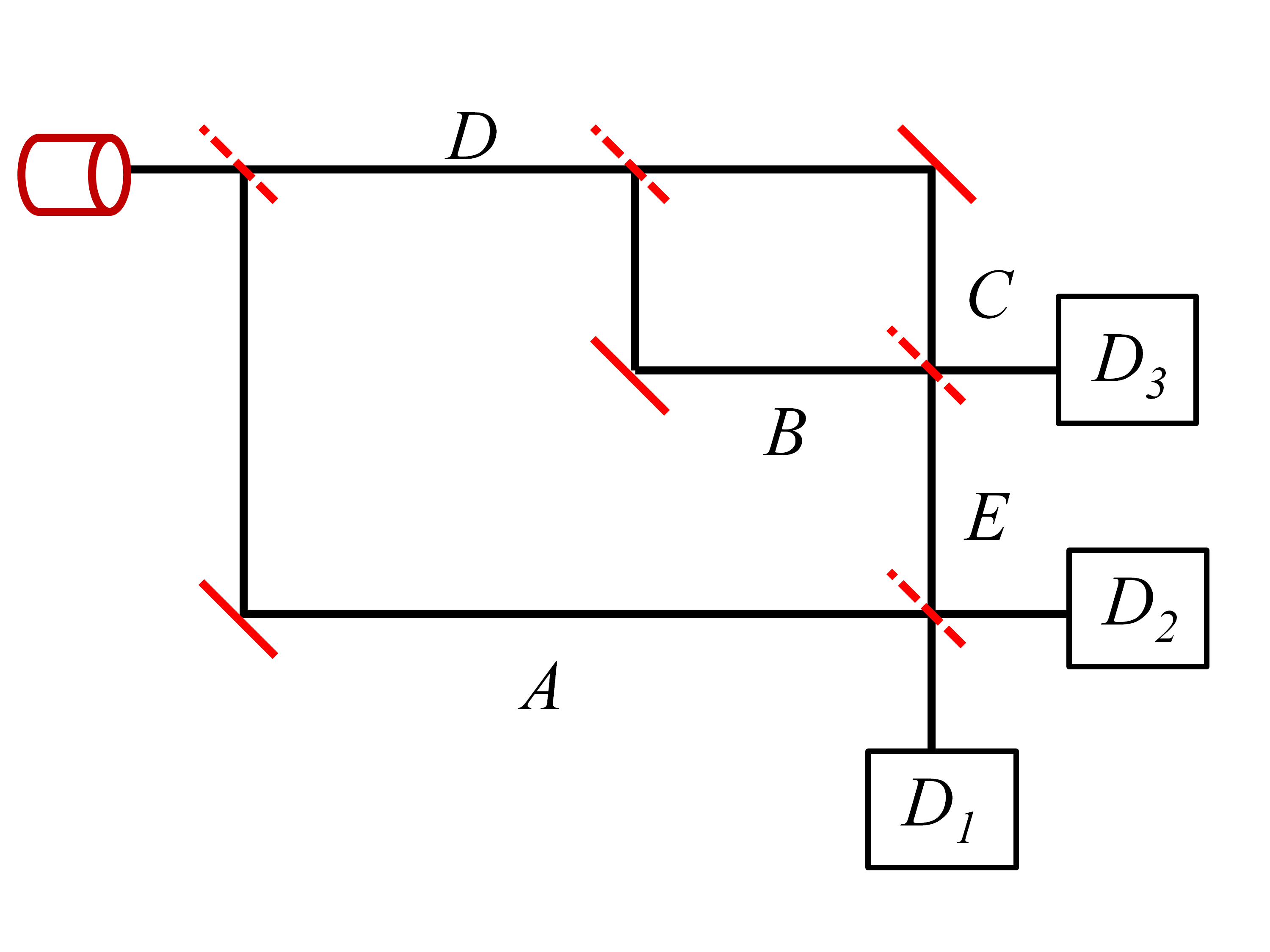}
	\caption{Vaidman's nested Mach-Zehnder interferometer setup.}
	\label{nestedmzi}
\end{figure}
In the nested Mach-Zehnder interferometer setup (Fig. \ref{nestedmzi}) considered in Vaidman's papers  \cite{pastofaparticle,tracing_the_past}, he argues that  whenever the particle is detected in the detector $D_{2}$ there is secondary presence in the branches $D$ and $E$.

The situation we considered in the second example above looks similar to the secondary presence in $D$ at first glance. The weak traces in branches $B$ and $C$ cancel each other, in a similar way to our example, where the traces of the pair cancel each other and give zero single-particle trace. However, we emphasize that it is different from the notion of secondary presence since there is already a primary presence in the form of non-vanishing two-particle traces in the boxes. As noted in the example above, the presence of the pair together in box 1 and 2 is also verifiable by a strong measurement scheme and, as we shall see in Section \ref{order}, the  two-particle traces can be of the same order as the single-particle traces depending upon the method of measurement used.

Next we consider a dynamic example similar to the second example.

\section{\label{dynamicex}A dynamic example}
Aharonov {\it et al.} considered a case of disappearing (and re-appearing) particle in \cite{eli}, and analyzed the particle's position at intermediate times between the postselection and preselection. Furthermore, an experimental scheme based on photonic quantum routers was suggested for testing this effect \cite{eli2}. We here analyze a variation  of the experiment and consider a pair of entangled particles with the preselection (at time $t=0$)

\begin{equation}
|\psi\rangle = \frac{1}{\sqrt{3}}\left(|11\rangle + i|22\rangle +|33\rangle\right),
\end{equation}
and postselection (at time $\ t = \frac{\pi\hbar}{\epsilon}$)
\begin{equation}
\langle \phi|= \frac{1}{\sqrt{3}} \left(-\langle 11|-i\langle 22| + \langle 33| \right).
\end{equation}

In what follows we will consider the case where only the first particle evolves under the Hamiltonian  $H=\epsilon\sigma_x$. At time $ t=\frac{\pi \hbar}{4\epsilon}$, the wavefunctions evolving from the past and future are (see Appendix for an auxiliary calculation):
\begin{equation}
|\psi\rangle = \frac{1}{\sqrt{3}} \left[
\frac{1}{\sqrt{2}}|11\rangle + \frac{1}{\sqrt{2}}|12\rangle - \frac{i}{\sqrt{2}}|21\rangle + \frac{i}{\sqrt{2}}|22\rangle + |33\rangle
\right]
\end{equation}

\begin{equation}
\langle\phi| = \frac{1}{\sqrt{3}} \left[
\frac{1}{\sqrt{2}}\langle 11| - \frac{1}{\sqrt{2}}\langle 12| + \frac{i}{\sqrt{2}}\langle 21| +\frac{i}{\sqrt{2}}\langle 22| + \langle 33|
\right].
\end{equation}

\begin{figure}
	\includegraphics[width=\linewidth]{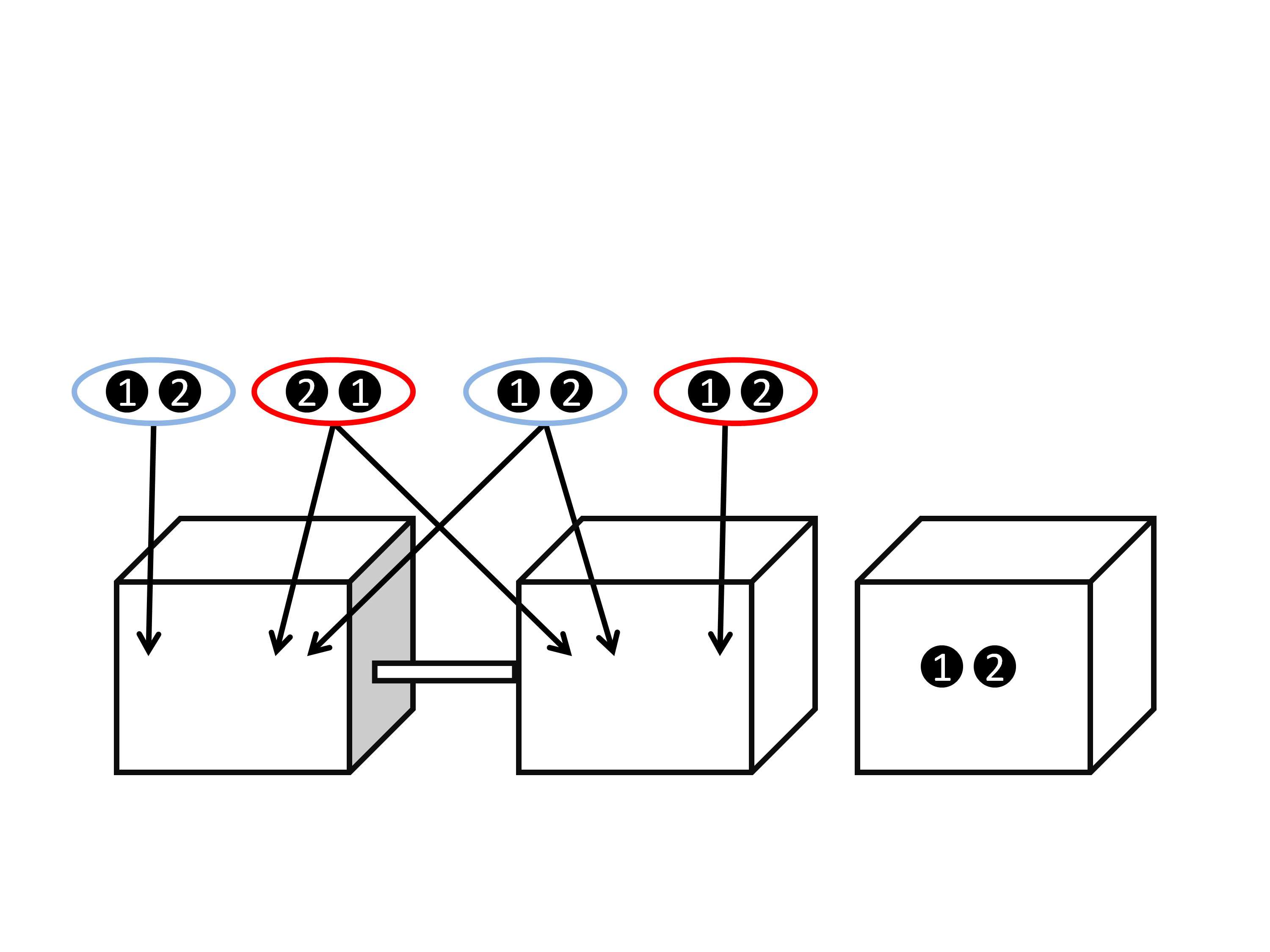}
	\caption{Two-particle correlations at time $t=\frac{\pi \hbar}{4\epsilon}$ for the modified disappearing (and re-reappearing) set-up. Blue oval denotes positive correlation while red denotes negative correlation.}
	\label{disappearing}
\end{figure}

At this instance, the weak values of the projection operators are

\begin{equation}
\begin{aligned}
 \langle P_1^{(1)}\rangle_{w}&=0, & \langle P_2^{(1)}\rangle_{w}&=0, \\
\langle P_{11}\rangle_{w}&=1/2, & \langle P_{12}\rangle_{w}&=-1/2, \\
\langle P_{21}\rangle_{w} &=1/2, &
\langle P_{22}\rangle_{w}&=-1/2.
\end{aligned}
\end{equation}
In the single-particle case considered in \cite{eli}, the first particle had disappeared from the first and second boxes and appeared in the third box. In this case too we
 see that the traces of the single-particle projections  into the first and second boxes are zero. However, similarly to the second example, there is a non-zero presence of pairs which is self-canceling Hence, we see again that considering only the single-particle traces would lead us to conclude that the first particle is absent from the first and second boxes. However, the non-zero pair traces tell us that there is nevertheless a subtle presence within these boxes.

 \section*{\label{nparticle}$N$\lowercase{-particle} C\lowercase{ase}}
 A more general case of $N$ particles in three boxes can be considered in a way similar to that of \cite{top-down} . Let the particles be prepared at the initial time $t=t_{i}$ in the state (without the overall normalizing factor),
 \begin{equation}
 |\psi\rangle = \prod_{i=1}^{N}\left(|1\rangle_{i}+|2\rangle_{i}\right) + \prod_{i=1}^{N}|3\rangle_{i}.
 \end{equation}
At a later time $ t=t_{f}$ the particles are postselected in
\begin{equation}
|\phi\rangle = \prod_{i=1}^{N}\left(|1\rangle_{i}-|2\rangle_{i}\right) + c\prod_{i=1}^{N}|3\rangle_{i},
\end{equation}
where $c$ is a constant.

 At any intermediate time $t_{i}<t<t_{f}$,  the traces of projection operators for the first and second boxes of 1 to $N-1$ particles considered together are zero. However, the $N$-th order traces are not zero. In other words, in the first and second boxes whenever we look for up to $N-1$ particles we do not find them there, but when we consider all the $N$ particles together we find a non-zero presence, i.e.,
\begin{eqnarray}
\begin{aligned}
\langle \prod_{i}^{N}\left(|1\rangle \langle 1| ~\text{or}~ |2\rangle \langle 2|\right)^{a_{i}}\rangle_{w} = 0,  \quad   \exists i.~a_{i} = 0,   \\
\langle \prod_{i}^{N}\left(|1\rangle \langle 1| ~\text{or}~ |2\rangle \langle 2|\right)^{a_{i}}\rangle_{w} =  \frac{{(-1)}^{n}}{c}, \quad \forall i.~a_{i}=1,
\end{aligned}
\end{eqnarray}
where $a_i$ are binary variables and $n$ refers to the number of projections into box 2.

As discussed previously, due to a property of a dichotomic operator \cite{dichotomic}, the zero correlations up to order $N-1$ can be corroborated by strong measurements. And for the same reason, when $c=1$ and $n$ is even, or when $c=-1$ and $n$ is odd, also the $N$-partite correlations can be verified via strong measurements.
 Furthermore, similarly to the discussion in Section \ref{simpleex}, all the correlations up to order $N-1$ can be obtained from the complete set of the higher $N$-partite correlations.

\section{\label{order} Strength of the weak trace of product operators}
Nonlocal or joint observables such as  $P_{11}$ and $P_{12}$ are difficult to measure directly as it is challenging to couple the measurement pointer to nonlocal operators without violating relativistic causality. Obtaining the weak trace of the joint observable sequentially by coupling to two pointers \cite{ReStein} gives us a trace proportional to the square of the coupling strength $g$. Since the single-particle traces are of first order in $g$, they might seem more significant than the traces of the joint operators (similarly to the difference between primary and secondary presence). However, if we use the Quantum Erasure method to perform non-local measurements as suggested in \cite{nonlocal_erasure}, the trace is proportional to $g$. The traces of the joint operators seem to be dependent on the choice of our method of measurement and are not necessarily weaker than the single-particle traces. This further reinforces our claim that multipartite correlations are just as important in determining the past of a particle.

\section{\label{conclusion} Conclusion}
The examples considered here show that the information  regarding the past of a particle in an entangled state obtained from the single-particle traces has to be augmented by the multi-particle traces. The trajectory of the pair, and in the case of more particles the higher order traces, also provide additional information about the past question. The original criterion for determining the past of quantum systems was motivated by whether there is any observable effect when we weakly couple a measurement pointer to the particle in a certain position. Therefore, we have claimed above that any kind of non-zero traces, either a single-particle trace or a multi-particle trace, should be considered as ``presence'' and should play a role when determining the past of a composite system. Another example emphasizing the importance of multipartite correlations was discussed in \cite{IFE}.

As explained in Section IV, the magnitude of the weak trace also depends on the measurement method and therefore traces of higher order projection operators are not necessarily smaller than the lower order traces. This supports our belief that any multipartite correlations cannot be ignored when trying to determine the whereabouts of an entangled particle. As described in Section \ref{simpleex} and \ref{dynamicex}, this view accords well with the recently proposed top-down \cite{top-down} hierarchical structure in quantum mechanics. We cannot construct the complete information about the past of a system solely from its single-particle traces. However, as all of the lower order traces, including single-particle ones, can be obtained from the complete set of higher order multi-particle traces, the higher order traces  seem to be essential for assessing the past of the particle in a more complete and consistent way.

\section*{\label{ack} Acknowledgements}
E.C. was supported by ERC AdG NLST.

\clearpage

\section*{Appendix - Calculation of the wavefunction evolution}
The preselection at $t=0$ and postselection at $t = t_f=\frac{\pi\hbar}{\epsilon}$ are:

\begin{align*}
|\psi\rangle = \frac{1}{\sqrt{3}}\left(|11\rangle + i|22\rangle +|33\rangle\right), \\
\langle \phi|= \frac{1}{\sqrt{3}} \left(-\langle 11|-i\langle 22| + \langle 33| \right)
\end{align*}

Since only the first particle evolves under $ H =\epsilon \sigma_x$ the evolution is given by
\begin{align}
|\psi(t)\rangle = U(t)\otimes I|\psi\rangle, \\
\langle \phi(t')| = \langle \phi|U^{\dagger}(t')\otimes I
\end{align}
where,
\begin{equation*}
t'= t_{f} - t,
\end{equation*}

\begin{equation*}
U(t)=e^{-\frac{i}{\hbar}\epsilon\sigma_xt}=
\begin{bmatrix}
\cos{\frac{\epsilon t}{\hbar}} & -i\sin{\frac{\epsilon t}{\hbar}} & 0 \\
-i\sin{\frac{\epsilon t}{\hbar}} & \cos{\frac{\epsilon t}{\hbar}} & 0 \\
0 & 0 & 1
\end{bmatrix},
\end{equation*}
and
\begin{equation*}
U^{\dagger}(t')=e^{-\frac{i}{\hbar} \epsilon\sigma_x t'}=
\begin{bmatrix}
\cos{\frac{\epsilon t'}{\hbar}} & -i\sin{\frac{\epsilon t'}{\hbar}} & 0 \\
-i\sin{\frac{\epsilon t'}{\hbar}} & \cos{\frac{\epsilon t'}{\hbar}} & 0 \\
0 & 0 & 1
\end{bmatrix}.
\end{equation*}
The two-state vectors at intermediate times are given by,

\begin{equation}
\begin{array}{lcl}
|\psi(t)\rangle = \\
 \frac{1}{\sqrt{3}} \left[\cos{\frac{\epsilon t}{\hbar}}  |11\rangle - i\sin{\frac{\epsilon t}{\hbar}}|21\rangle
 	   + \sin{\frac{\epsilon t}{\hbar}}|12\rangle + i\cos{\frac{\epsilon t}{\hbar}}|22\rangle
 	   + |33\rangle \right]
\end{array}
\end{equation}

and

\begin{equation}
\begin{array}{lcl}
\langle\phi(t')| = \\
 \frac{1}{\sqrt{3}} \left[
-\cos \frac{\epsilon t'}{\hbar} \langle 11| + i\sin\frac{\epsilon t'}{\hbar}\langle 21| - \sin\frac{\epsilon t'}{\hbar}\langle 12| - i\cos\frac{\epsilon t'}{\hbar}\langle 22| + \langle 33|
\right]
\end{array}
\end{equation}

\end{document}